# Oxygenation and Blood Volume Periodic Waveforms in the Brain

Alexander Gersten,[1,2] Dov Heimer[3], Amir Raz[4]

[1] Department of Physics, Ben-Gurion University, Beer-Sheva, Israel
[2] Department of Psychology, Hunter College of the City University of New York, NY, USA
[3] Pulmonary Unit, Soroka University Hospital, Beer-Sheva, Israel
[4] Columbia University College of Physicians & Surgeons, New York, NY USA

e-mail: alex.gersten@gmail.com

## Abstract

Results of an experiment are presented whose aim is to explore the relationship between respiration and cerebral oxygenation. Measurements of end tidal $CO_2$ (EtCO$_2$) were taken simultaneously with cerebral oxygen saturation (rSO$_2$) using the INVOS Cerebral Oximeter of Somanetics. Due to the device limitations we could explore only subjects who could perform with a breathing rate of around 2/min or less. Six subjects were used who were experienced in yoga breathing techniques. They performed an identical periodic breathing exercise including periodicity of about 2/min. The results of all six subjects clearly show a periodic change of cerebral oxygenation with the same period as the breathing exercises. Similar periodic changes in blood volume index were observed as well.

_______________________________________________________________________

**INTRODUCTION**

Breathing may have dramatic effects on the human cerebral blood flow (CBF) and cognition. This was already known long time ago to Chinese (Li Xiuling, 2003), Hindus (Kuvalayananda, 1983) and Tibetans (Mullin, 1996).

The main parameter influencing the CBF is the arterial partial pressure of carbon dioxide ($PaCO_2$). About 70% increase in $PaCO_2$ may double the blood flow (Sokoloff, 1989, Guyton, 1991). The increase of blood flow to the brain results in an increase of nutrients and oxygen, which may influence to great extent brain's physiology.

An evaluation of the CBF dependence on $PaCO_2$ was given by Gersten (Gersten 1999). The human normal value of $PaCO_2$ is about 40 mmHg. Fig. 1 depicts our estimation of the changes of CBF as a result of changing $PaCO_2$ from normal. The estimate is based on the data of Refs. (Reivich 1964, Ketty and Schmidt 1948, Raichle et all 1970). An increase of only 12.5% from normal in $PaCO_2$ leads to the state of hypercapnia ($PaCO_2$ above 45 mm Hg).

With aging both CBF and $PaCO_2$ tends to fall down, therefore breathing exercises (or procedures) with the aim to increase $PaCO_2$ and CBF might be important in preventing and alleviating neurodegenerative diseases. They may be also important in improving intellectual abilities.

The amount of $PaCO_2$ in our arteries is

1. Proportional to the $CO_2$ which was metabolized by the organism.
2. Inversely proportional to the lung ventilation (i.e. to the amount of air exchanged between the lungs and the environment in one minute).

An instantaneous way to change $PaCO_2$ can be accomplished by influencing the ventilation.

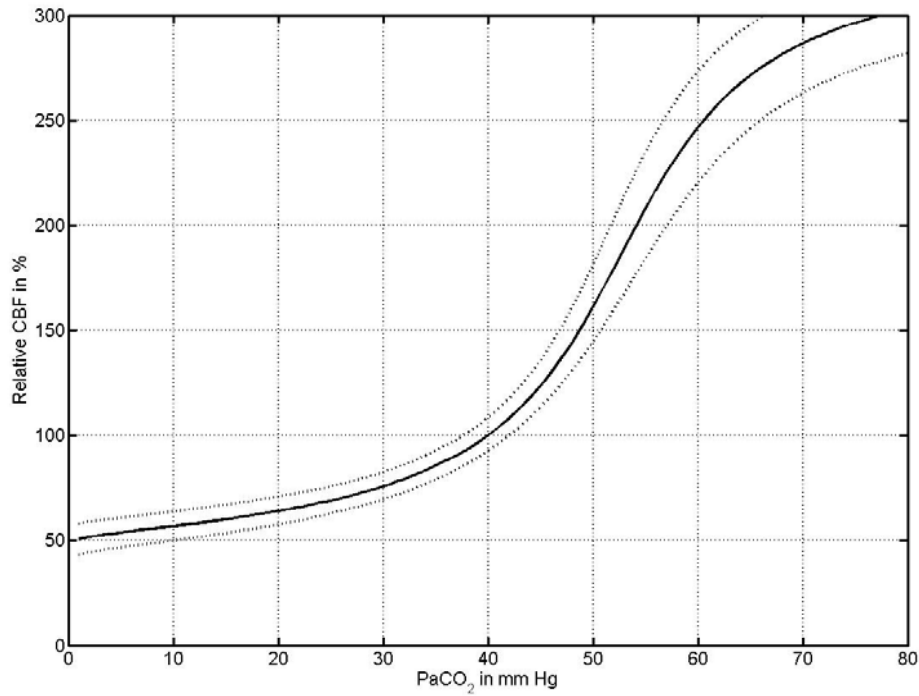

**Fig. 1. Estimated changes of human CBF from normal (100%) against $PaCO_2$.
Normal $PaCO_2$ =40 mm Hg is assumed, it corresponds to ordinate value of 100%.**

Respiratory periodicities were observed in fMRI, but they were treated as artifacts (Windischberger et al., 2002) or as a noise (Raj et al., 2001). The main research effort was directed towards the elimination of the effect of the respiratory periodicities. Our aim is to study the effect of respiration on brain's oxygenation and performance.

Near infrared spectroscopy (NIRS) is an efficient way to probe brain's oxygenation (Rolfe, 2000, Thavasothy et al., 2002). Brain oxygenation increases with an increase of $PaCO_2$ (Imray et al., Thavasothy et al., 2002). In this paper we examine the dependence of brain's oxygenation on $PaCO_2$.

**METHODS AND MATERIALS**

We started this research in Israel in the Pulmonary Unit of the Soroka University Hospital in Beer-Sheva. The study protocol was approved by the Helsinki (Ethics) Committee of the Soroka University Hospital. The investigation conforms with the principles outlined in the Declaration of Helsinki. The nature of the study was explained, and all subjects gave written consent to participate.

At the beginning our research was based on capnometer measurements of end tidal $CO_2$ ($EtCO_2$). The capnometer measures the $CO_2$ concentration of the expired air. During the inspiration or breath holding the capnometer indications are zero. The capnometer enable us to follow the breathing periodicity.

We continued the research in the USA using the INVOS Cerebral Oximeter model 5100B of Somanetics Corporation and a capnometer of Better Physiology, Ltd. Both devices are non-invasive. The INVOS Cerebral Oximeter is based on most recent technological developments of near infrared spectroscopy (NIRS). With this device data are collected of regional oxygen saturation ($rSO_2$) near the forehead with two optical sensors (for more details see www.somanetics.com). In addition to $rSO_2$ one can determine the Blood Volume Index (BVI), which is an indicator of blood changes in the brain. This is a relative quantity, which could not be normalized with our oximeter .

The fastest recording rate of the 5100B oximeter is every 12 seconds from both left and right sensors. This time is much longer then the average period of about 4 seconds of normal respiration. In order to detect oxygenation periodicity we had to study respiration periods of about 36 seconds or larger (i.e. 3 data points or more for each breathing period). This is still a rather small amount of data points per respiration period. We have compensated for this small number by using a cubic spline interpolation of the data points, adding new interpolation points through this method. The cubic spline interpolation is a very effective method of smooth interpolation.

We found six people well acquainted with yoga pranayama, who could easily perform breathing exercises with periods around 36 seconds. All of them performed the following routine which lasted for 15 minutes.

They were asked to breathe in the following way: to inhale for 4 units of time (UOT), to hold the breath for 16 UOT and to exhale for 8 UOT, this we denote as the 4:16:8 (pranayama) routine.

The unit of time (UOT) is about 1 second. The yoga practitioners develop an internal feeling of UOT which they employ in their practices. They learn to feel their pulse or they learn to count in a constant pace. Often they practice with eyes closed. In order not to distract or induce additional stress we preferred not to supply an external uniform UOT. The primary concern for this research was to have a constant periodicity and in this case the practitioners have succeeded to maintain it. The data were analyzed with spectral analysis which took into account non-stationary developments, which were subtracted from $rSO_2$ and BVI data. Sharp picks corresponding to the breathing periodicity were found in the spectral analysis of $EtCO_2$ (the amount of $CO_2$ during expiration).

The motivation for this exercise was to see the relation of $rSO_2$ (oxygenation) to the increasing amount of $PaCO_2$. Actually instead of $PaCO_2$ the end tidal CO2 (EtCO2) was measured (the maximal $CO_2$ at exhalation), a quantity which approximate well the $PaCO_2$. The periodicity of rSO2 was studied during the 4:16:8 respiration period (of approximately 32 UOT).

The $rSO_2$ was measured with the aid of two sensors placed on the forehead, detecting oxygenation from the left hemisphere (frontal part in about 3 cm depth) and the right hemisphere (frontal part in about 3 cm depth). At the same time data for evaluation of BVI were collected.

**RESULTS**

In Fig. 2 the results of simultaneous measurements of $EtCO_2$, $rSO_2$ and differences in BVI are presented for the 6 subjects.

In order to check more precisely the periodic behavior, spectral analysis was performed on the $EtCO_2$ data and the interpolated $rSO_2$ and BVI data. The $rSO_2$ and BVI data have a non-stationary component. The spectral analysis was performed on the raw $EtCO_2$ data and on $rSO_2$ and BVI interpolated data from which the non-stationary components were subtracted. The results are shown in Fig. 2. and Table 1. There is a clear overlapping between the periodicity of $EtCO_2$ and the periodicities of $rSO_2$ and BVI.

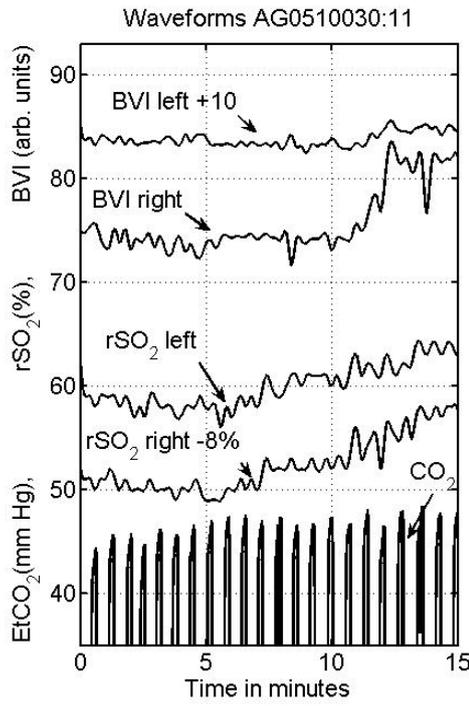
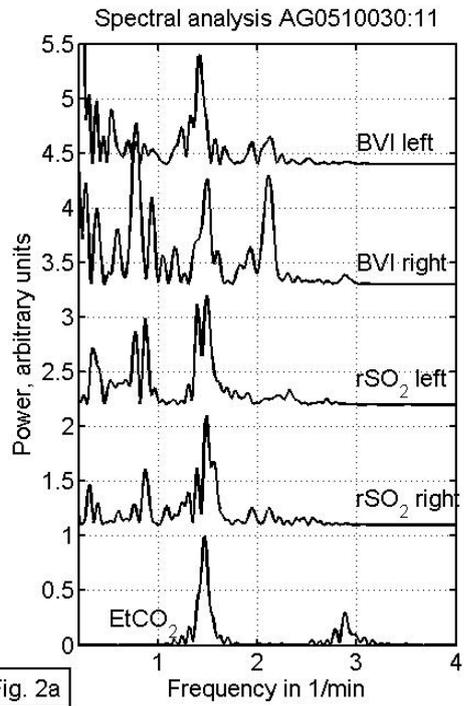
Fig. 2a

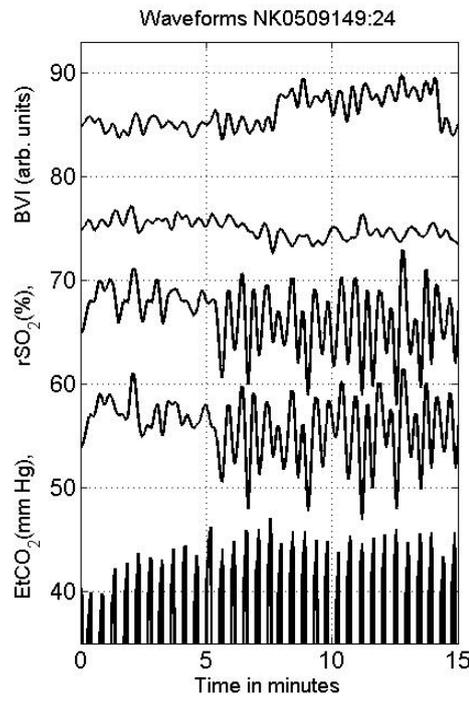
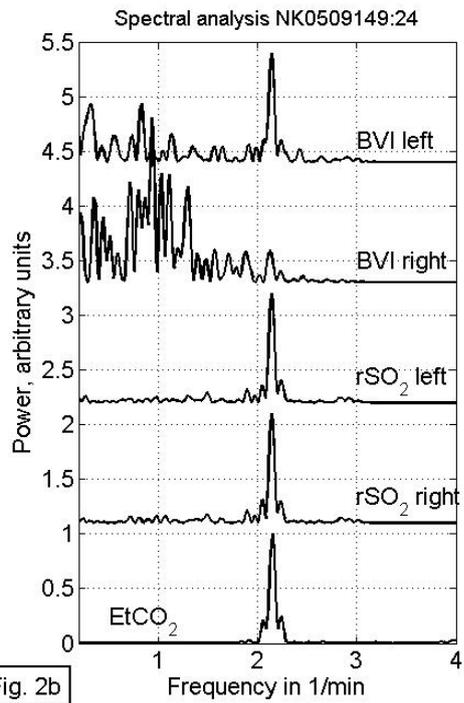
Fig. 2b

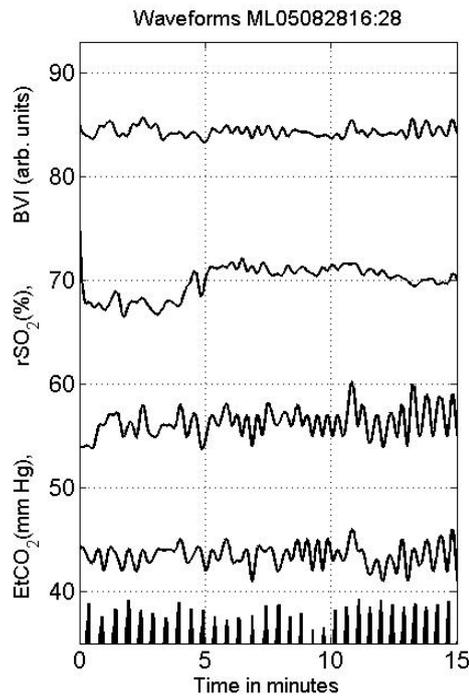
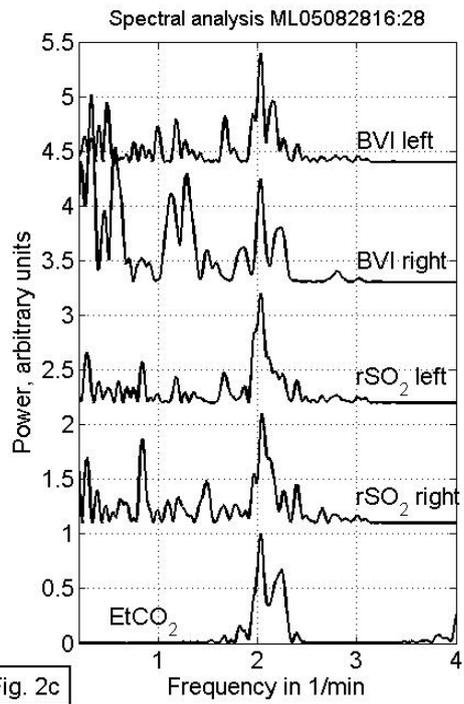

Fig. 2c

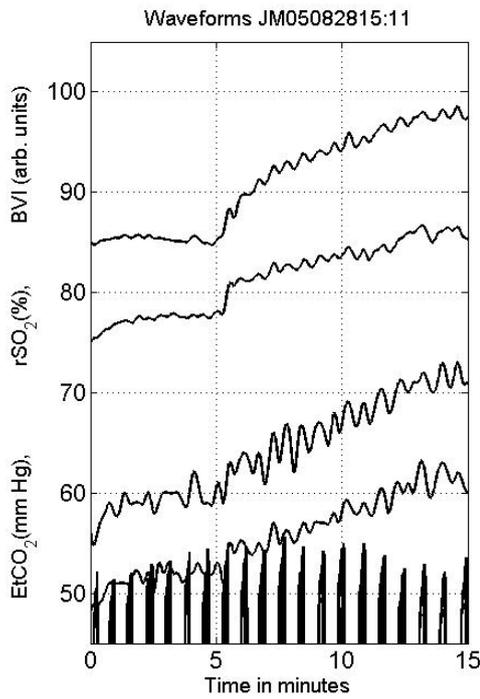
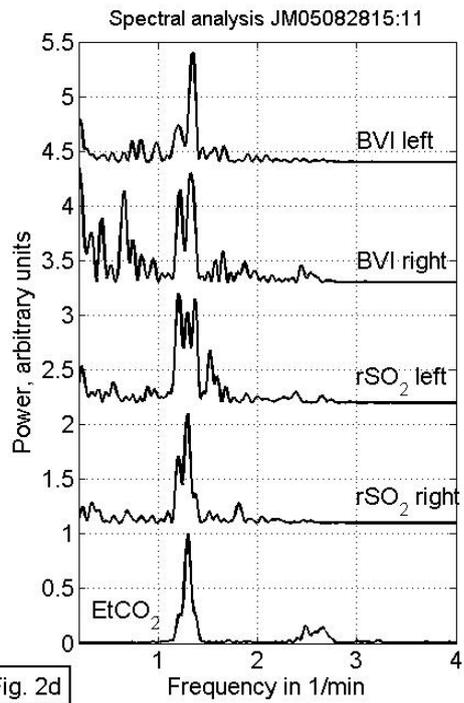

Fig. 2d

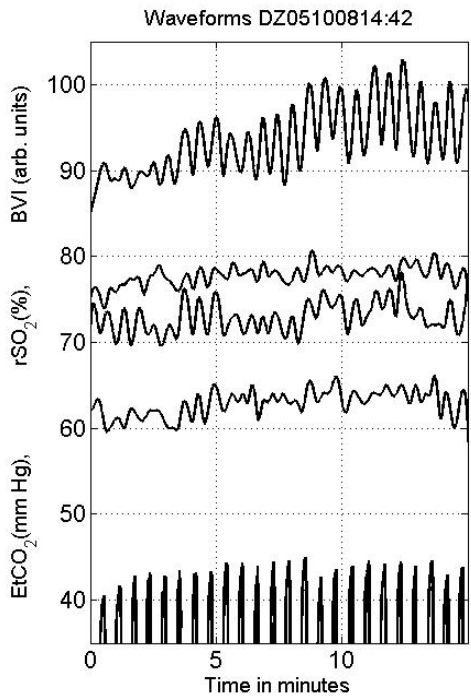
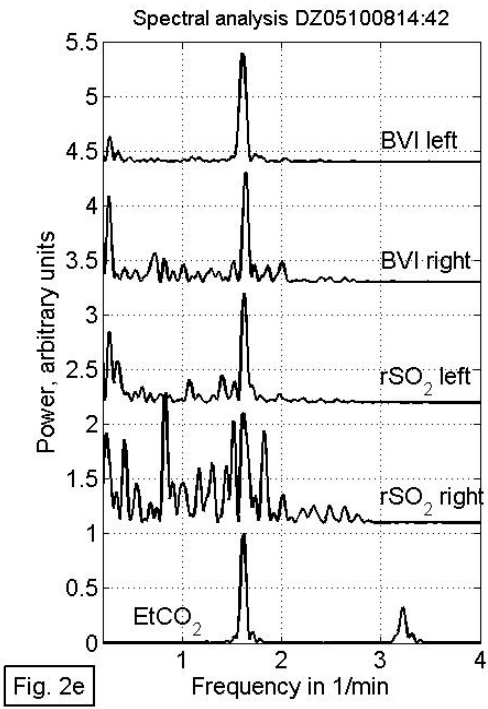

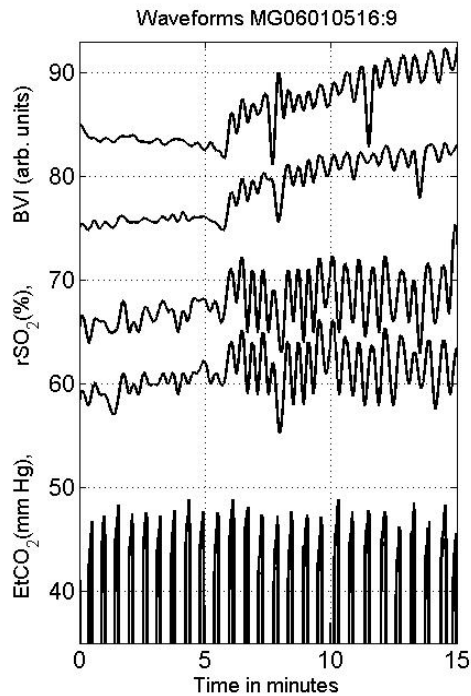
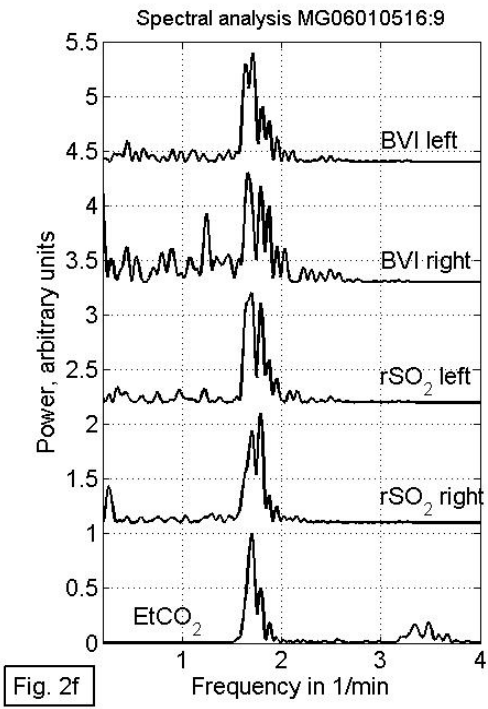

**Fig.2** On the left hand side are given from the bottom to top: the readings of the capnometer in mm Hg, rSO₂ from the right sensor (subtracted with 8%), rSO₂ from the left sensor, the difference of BVI from the right sensor, and the difference of BVI from the left

**sensor (increased by 10). On the right hand side the corresponding spectral analyses of the waveforms are given.**

**Table 1**

The position of the dominant frequencies (in units of 1/minute) of Fig. 2 in the spectral analysis, in parenthesis the extension of the half width is given.

|    | PaCO$_2$ | rSO$_2$ right | rSO$_2$ left | BVI right | BVI left |
|----|----------|---------------|--------------|-----------|----------|
| **AG** | 1.47 (1.42-1.51) | 1.49 (1.45-1.53) | 1.48 (1.45-1.54) | 1.49 (1.42-1.53) | 1.42 (1.38-1.47) |
| **NK** | 2.15 (2.12-2.18) | 2.14 (2.12-2.17) | 2.14 (2.12-2.17) | ------------------ | 2.14 (2.11-2.17) |
| **ML** | 2.03 (1.98-2.07) | 2.05 (2.01-2.15) | 2.03 (1.95-2.08) | 2.03 (1.99-2.07) | 2.03 (2.00-2.06) |
| **JM** | 1.30 (1.26-1.34) | 1.30 (1.26-1.33) | 1.30 (1.17-1.40) | 1.33 (1.29-1.37) | 1.35 (1.31-1.38) |
| **DZ** | 1.62 (1.59-1.65) | 1.62 (1.59-1.67) | 1.62 (1.60-1.65) | 1.63 (1.61-1.67) | 1.61 (1.57-1.65) |
| **DG** | 1.70 (1.65-1.73) | 1.79 (1.65-1.82) | 1.70 (1.62-1.82) | 1.66 (1.62-1.89) | 1.71 (1.61-1.74) |

**DISCUSSION**

We have verified in this experiment that breathing may affect to great extent the brain physiology. With breathing exercises the arterial $CO_2$ can be increased, which in turn increases CBF and brain oxygenation.

The most important finding of this experiment is the periodic correlation between respiration, oxygenation and blood volume changes. The results of all six subjects, as shown in Table 1 and Fig. 2, clearly show a periodic change of cerebral oxygenation with the same period as the breathing exercises, indicating that with each breath the brain oxygenation was periodically changing. Similar periodic changes in blood volume indicate that the brain pulsates with a frequency of respiration.

Present results were achieved by using very slow breathing patterns. If the present devices will be modified to allow sampling of rSO$_2$ at frequencies higher than the frequency of normal respiration and higher than the heart rate, it will be possible to observe new types of brain waveforms. These waveforms may have new information about brain oxygenation, cognitive function, brain pulsation and brain motion. Under these circumstances it will be possible to understand the correlations between respiration and brain physiology much better. The sampling of rSO$_2$ should go up to or preferably above 4 Hz. The accuracy of the reading device also should be changed from 2 significant digits to 3 digits. We believe that these changes will enable new explorations and new insights on the influence of respiration on brain's physiology. Neurodegenerative diseases are characterized by low CBF, in our research we have found effective ways to increase CBF. We will continue our research in order to explore the possibilities of increased CBF and its influences on intellectual abilities and on fighting degenerative diseases. Devices displaying the oxygenation periodic waveforms should be developed for new diagnostic and research purposes.

**Acknowledgements**

We are grateful to Tom O'Flaherty, Jnr., Northeast Regional Manager, Somanetics Corporation for his encouragement and support, without which this research could not have been done.

We thank Leslie Kaminoff and Nigol Koulajian for their encouragement and help to find volunteers.